**Timing and Dynamics of the Rosanna Shuffle**

Esa Räsänen[1], Niko Gullsten[1] Otto Pulkkinen[2], Tuomas Virtanen[3]

[1] Computational Physics Laboratory, Tampere University, Finland

[2] Department of Mathematics and Statistics, University of Turku, Finland

[3] Audio Research Group, Tampere University, Finland

**Author Note**

**Abstract**

The Rosanna shuffle, the drum pattern from Toto's 1982 hit "Rosanna", is one of the most recognized drum beats in popular music. Recorded by Jeff Porcaro, this drum beat features a half-time shuffle with rapid triplets on the hi-hat and snare drum. In this analysis, we examine the timing and dynamics of the original drum track, focusing on rhythmic variations such as swing factor, microtiming deviations, tempo drift, and the overall dynamics of the hi-hat pattern. Our findings reveal that "Rosanna" exhibits a surprisingly pronounced swing for its genre, along with notable tempo drift typical of tracks recorded without a metronome. Additionally, we observe clear long-range correlations in the microtiming deviations, consistent with previous studies. Notably, the two-bar phrases of the song feature a distinctive repeating pattern in the timing and dynamics of the hi-hat beats, which enhances the song's phrasing. Overall, the Rosanna shuffle boasts a rich array of rhythmic characteristics that solidify its significant place in music history.

**Introduction**

When humans perform rhythms, whether through music, speech, or movement, they often deviate from the precise timings specified in musical notation or metronomic instructions. These deviations, rather than being seen as errors, are a natural part of human expression and have been the subject of extensive interdisciplinary research (Repp, 1999; London, 2012). Understanding these subtle variations offers insights into the nature of human cognition, motor control, and emotional expression, and contributes to the fields of psychology, musicology, neuroscience, and even mathematical modeling.

Human rhythmic deviations in performance are influenced by both cognitive and motor processes. Timing variations emerge from the brain's adaptable internal clocks, which respond to external cues and internal physiological states, enabling performers to synchronize with a beat (Large & Palmer, 2002). These deviations, whether intentional (like rubato) or unintentional, enhance the expressiveness and human quality of the performance (Keller, 2020). Moreover, natural fluctuations, referred to as microtiming deviations, occur due to limitations in motor control and cognitive factors (Repp, 2005).

Microtiming deviations in music range from a few to several tens of milliseconds relative to the precise beat and often exhibit long-range correlations (LRC), displaying fractal-like patterns similar to 1/f noise (Hennig, 2012). Such timing fluctuations are not only observed in laboratory settings (Hennig, 2011) but also in recorded pop and rock music (Räsänen, 2015). These timing variations reflect underlying neurobiological processes, which can be affected in conditions like Parkinson's disease but may be restored through interventions such as deep brain stimulation (Ruiz, 2014). Ongoing studies in auditory processing are exploring how the brain perceives and reacts to these microtiming variations (Doelling & Poeppel, 2015).

From a listener's viewpoint, macro- and microscale timing deviations are often seen not as errors but as elements that enhance the richness and complexity of a performance. Research in music perception indicates that listeners are highly attuned to timing variations, often favoring performances with slight fluctuations over rigid, mechanical rhythms (Large & Palmer, 2002). However, preferences can vary based on musical style and the listener's level of expertise (Davies, 2013). Studies have shown that subtle timing deviations, especially in genres like jazz, are crucial for creating groove and feel (Butterfield, 2011). Yet, in the case of swing feel, it has been observed that naturally occurring microtiming deviations are not necessarily essential. Regardless, these rhythmic intricacies influence the listener's emotional and physical reactions to music, fostering a deeper connection to the performance. A recent review of psychological and neuroscientific research on musical groove underscores these points (Etani, 2024).

In this study, we examine the timing and dynamics variations of the iconic half-time shuffle drumbeat from *Rosanna*, performed by Jeff Porcaro. This groove has become a quintessential example of advanced drumming in popular music and is frequently analyzed outside a scientific context for its technical intricacies and rhythmic feel (Glass, 2017). Porcaro's blend of precision and groove in this track has garnered attention in music production and drumming tutorials. However, to our knowledge, there have been no prior scientific investigations into the detailed timing and dynamics variations of this drum pattern. Our analysis focuses on the swing feel in the hi-hat pattern, tempo fluctuations, rhythmic structuring, and the distinctive patterns of timing and dynamics variations, including an examination of long-range correlations.

# Method

**Musical Recording**

"Rosanna" is the opening track on Toto's fourth studio album, *Toto IV*, released in 1982 (Toto, 1982). The song held the number two position on the Billboard Hot 100 for five consecutive weeks and won the Grammy Award for Record of the Year in 1983. Recently, the musical elements of "Rosanna" have been thoroughly analyzed by Rick Beato, a well-known YouTube personality, record producer, session musician, and university professor (Beato, 2018). The song features one of the most recognizable drumbeats in music history, known as the Rosanna shuffle, performed by Jeff Porcaro (1954-1992). Porcaro is one of the most recorded drummers ever and, along with the other members of Toto, one of the most prolific session musicians of the 1970s and 1980s.

Jeff Porcaro cited three records as the primary influences for the drumbeat: Steely Dan's "Babylon Sisters" and "Home at Last," both played by Bernard Purdie, and Led Zeppelin's "Fool in the Rain," performed by John Bonham (Corp, 2012). The Rosanna shuffle (Fig. 1) is made up of triplets, with the first and third notes played on the hi-hat and the middle note serving as a ghost-note on the snare. The third beat of each bar is accented with the snare, while the bass drum follows a Bo Diddley pattern, a variation of the well-known 3-2 son clave rhythm.

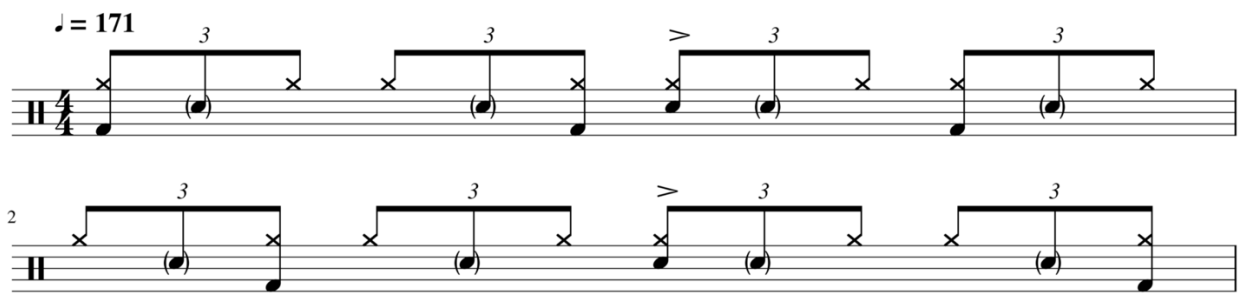

*Figure 1. Drum score for two bars of the Rosanna shuffle played in the intro and the verse.*

The Rosanna shuffle, illustrated in Fig. 1, occurs a total of 60 bars in the song: 12 bars for the intro and 48 bars for the verses. The chorus features a similar shuffle, but the bass drum follows a different pattern that accents the hits from the guitar, keyboard, and horns. Porcaro also incorporates several fills in the choruses, which influence the onset detection (see below). In the pre-chorus section, "Not quite a year since you went away…," the groove shifts to a straight quarter-note feel, as shown in Fig. 2.

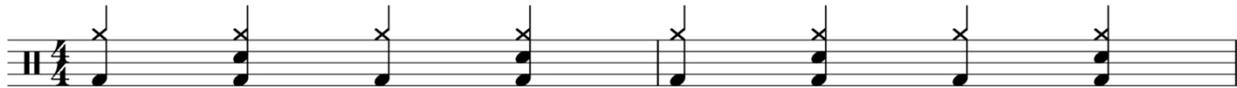

*Figure 2. Drum score for two bars of the pre-chorus section.*

**Onset Detection**

The extraction of hi-hat onsets from the raw drum track was performed similarly to a previous study on hi-hat drumming patterns by some of the current authors (Räsänen et al., 2015). The original raw drum track was processed using a high-pass filter with a cutoff frequency of 1 kHz. Automated onset detection was conducted by identifying local maxima in the amplitude envelope of the signal, where the amplitude exceeded a manually adjusted threshold. The automatic onset detection was followed by manual correction, where missed onsets were added, extra onsets removed, and the timings of the onsets were edited to visually match the attacks in the amplitude envelope as well as the perceived onset timings. Based on visualization of the signal envelope, the uncertainty in the timings of the onsets was estimated to be typically two milliseconds. For sharp drum strokes without interference from other cymbals or drums in the same frequency range, the uncertainties of the onset times were estimated to be less than one millisecond. We set a threshold of five milliseconds and discarded any onsets that exceeded this limit. There were several instances where multiple hits were detected 2-3 milliseconds apart for the same triplet; in these cases, the onset was annotated to the first hit. In addition to the timing of the hi-hat onsets, their amplitude (dynamics) was also recorded.

A total of 1239 hi-hat onsets were identified. The intervals that make up the Rosanna shuffle and the pre-chorus were extracted. Intervals exceeding 3.5 times the average interval within a triplet were considered to result from missing onsets and were therefore excluded from the analysis. The total number of detected intervals is 1154, which corresponds to a 93% detection rate.

**Numerical Concepts and Methods**

In our analysis, we consider the onsets of the hi-hats as a time series, with timestamps denoted as $f(i)$ and amplitudes denoted as $g(i)$. The interbeat intervals are calculated as $\tau(i) = f(i+1) - f(i)$. We can now define the *drift* of the tempo as (Räsänen, 2015)

$$d(i) = \sum_{j=1}^{i} \tau(j) - i \langle \tau \rangle,$$

where $\langle \tau \rangle$ is the mean duration of the single eight-note triplet. Thus, the drift compares the cumulative sum of the intervals at a given time to a temporal grid of an imaginary metronome. If an interval was found to be close to a multiple of two or three of a single eighth-note triplet, it was divided by two or three, respectively. If the interval exceeded a multiple of 3.5 of a single eighth-note triplet, it was discarded, and the drift was set to zero.

We also analyze the tempogram of the song generated using Sonic Visualiser software (Cannam, 2010) with the "Tempogram" Vamp plug-in developed by Carl Bussey, based on the work of Grosche (2010). The parameters used were as follows: Novelty curve spectrogram compression constant = -1000, Novelty curve minimum dB = -74, Tempogram window length = 1024, Tempogram hop size = 64, Tempogram FFT length = 4096, Cyclic tempogram minimum BPM = 30, Cyclic tempogram maximum BPM = 360, Cyclic tempogram octave divider = 60, and Cyclic tempogram reference tempo = 84.

In a shuffle drumbeat, the eighth-note hi-hat patterns are usually played in a long-short pattern, where the notes on the beat are extended to produce a shuffle feel. The ratio of the consecutive eighth notes is referred to as the swing ratio. Figure 3 illustrates the definition of the swing ratio for the Rosanna shuffle, indicating that, according to the score, the theoretical swing ratio is 2:1.

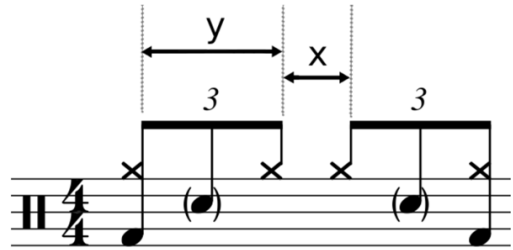

Figure 3. Swing ratio for the Rosanna shuffle is defined as y/x.

To enhance our statistical analysis of the interbeat intervals, we utilize detrended fluctuation analysis (DFA) (Peng, 1994; Peng, 1995; Kantelhardt, 2001) to investigate the correlations in the timing and dynamics of the Rosanna shuffle. DFA is a commonly used technique for analyzing nonstationary time series across various fields, from heartbeat dynamics (Peng, 1995) to drumming (Räsänen, 2015). For a comprehensive introduction to DFA and its recent developments, such as dynamical DFA, we refer to a recent paper by Molkkari et al. (Molkkari, 2020) and summarize only the key steps here:

1. The mean value of the (similar type of) interbeat intervals is subtracted from all the intervals.
2. A cumulative sum of the intervals obtained in step 1 is computed.
3. The data is split to sets of non-overlapping windows of same size. Below, the window size is referred to as the scale *s*.
4. For each scale, polynomials of order n, i.e., the trends are fitted to the cumulative sum of the intervals, and a detrended time series is computed as the difference. Here we use linear detrending with n=1. For each window (of each scale), the variance is computed.
5. For each scale, the variances are averaged over all the windows and a square root is taken to obtain the fluctuation function $F(s)$.

6. The scaling properties of the fluctuation function are examined through $F(s) \propto s^{\alpha}$, where α is the scaling exponent.

The value of the scaling exponent is linked to the characteristics of the data. Values $\alpha = 0.5$ and $\alpha = 1.5$ correspond to (uncorrelated) white noise and (strongly correlated) Brownian noise, respectively. On the other hand, $\alpha = 1$ corresponds to pink noise, flicker noise or 1/f noise, often referred to as "fractal" noise. Our focus here is on the range $0.5 < \alpha \leq 1$, which indicates the presence of *long-range correlations* in the time series. It is important to study the scaling exponent as a function of scale, as the data may exhibit very different characteristics at short and long scales. Typically, $\alpha$ is split into $\alpha_1$ in short scales $s = 4 \ldots 16$ and $\alpha_2$ in long scales $s = 16 \ldots 100$ (Kantelhardt, 2001). The analysis can be further refined by examining $\alpha(s)$ continuously across scales (Molkkari, 2018) and dynamically in the form of $\alpha(s,t)$ as a function of both scale and time (Molkkari, 2020).

## Results

**Classification and Statistics of the Intervals**

Figure 4 displays the raw interval (a) and amplitude data (b) extracted from the hi-hat onsets of Rosanna. The large peaks in Fig. 4(a) are due to the missing onsets and will be removed in the subsequent time-series analysis, as previously described. Aside from these peaks, the series exhibits a zig-zag pattern, as shown in the inset, indicating an alternating short-long interval pattern. This behavior primarily results from the absence of ghost notes in the triplets throughout most of the song, leading to an alternation between longer intra-triplet and shorter inter-triplet intervals. Occasionally, however, this pattern is disrupted by a detected ghost note on the second beat of the triplet, as illustrated in the middle of the inset in Fig. 4. Additionally, the three pre-chorus sections, characterized by straight drumming patterns and relatively long hi-hat intervals, are clearly visible, particularly around interval numbers close to 300.

For the intervals being analyzed, we define the terms *singles*, *doubles*, and *triples*, as illustrated in Figs. 5(a), (b), and (c), respectively. These names refer to the ratios of 1:2:3 between these intervals in musical notation. The singles and doubles form the Rosanna shuffle, with singles defined as the interval between the last and first notes of different triplets, while doubles are defined as the interval between the first and last notes of the same triplet. The triples consist of the consecutive intervals in the straight-beat pre-chorus of the song.

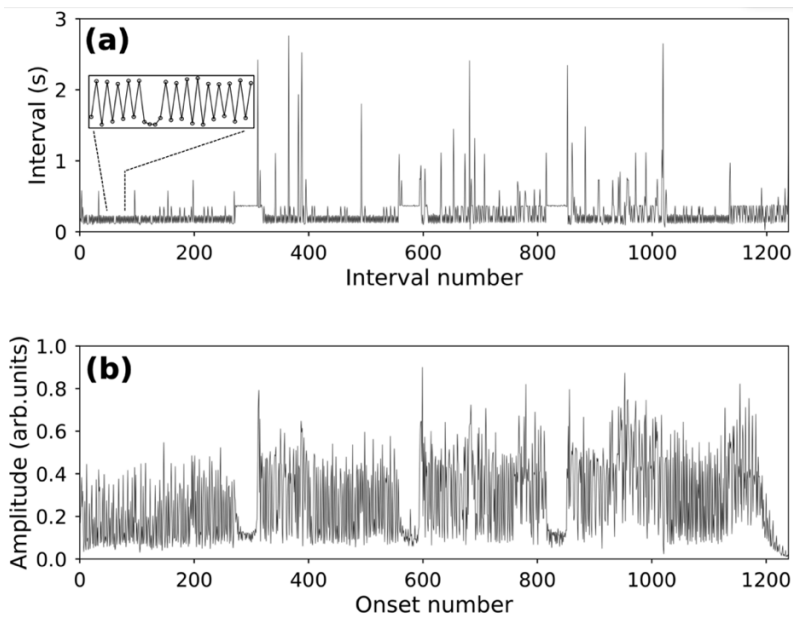

Figure 4. Raw data of the interbeat intervals extracted from the hi-hat onset times (a) and the amplitudes of the onsets (b).

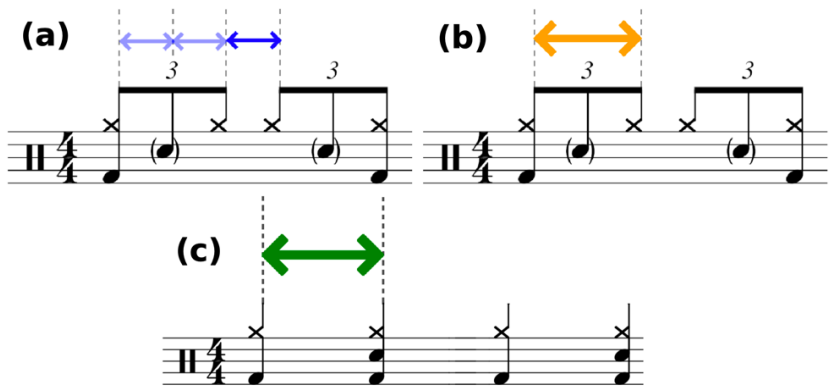

*Figure 5. Intervals under examination in the study, denoted in the text and in the following figures as singles (a), doubles (b) and triples (c), respectively.*

Theoretically, we would expect a roughly equal number of singles and doubles if we were only considering the hi-hat onsets. However, as mentioned earlier, occasional ghost notes on the snare are detected, particularly at the beginning of the song. We included the corresponding intra-triplet intervals (hi-hat – snare – hi-hat) as singles in our analysis.

A total of 454 singles, 421 doubles, and 279 triples were identified. These are shown in Fig. 6 as a function of the corresponding interval count *N*. The histograms for each interval class can be found in Fig. 7. The mean interval lengths and standard deviations for singles, doubles, and triples are 127.6 ± 8.4 ms, 228.7 ± 8.4 ms, and 364.0 ± 7.8 ms, respectively. As anticipated and consistent with previous studies (Hennig, 2015; Räsänen, 2016), the histograms generally exhibit Gaussian distributions. This pattern results from the effort to maintain a steady tempo throughout the song. However, as will be demonstrated below, there are additional dynamic effects in the interval variations that are not captured in the histograms, which average out the time-dependent aspects.

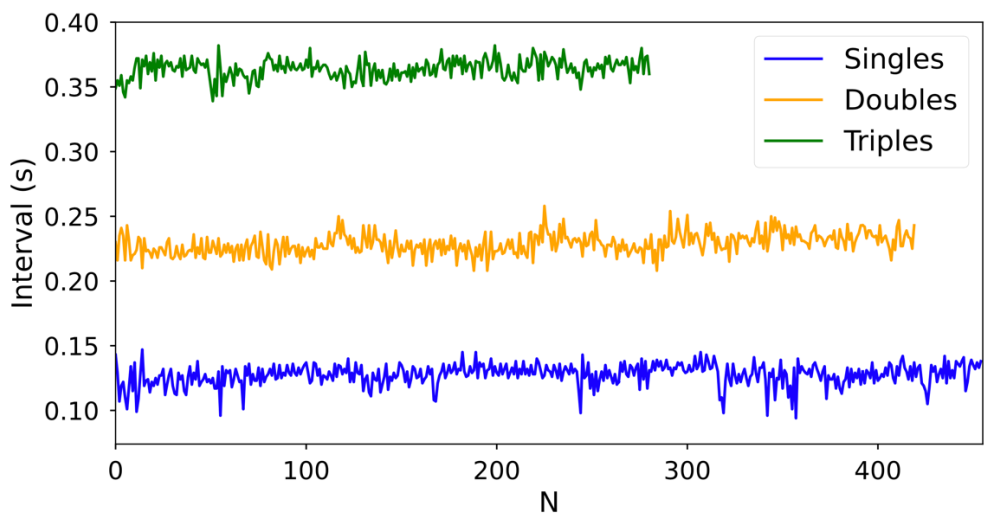

*Figure 6. Evolution of three types of intervals in "Rosanna" according to the notation in Fig. 5.*

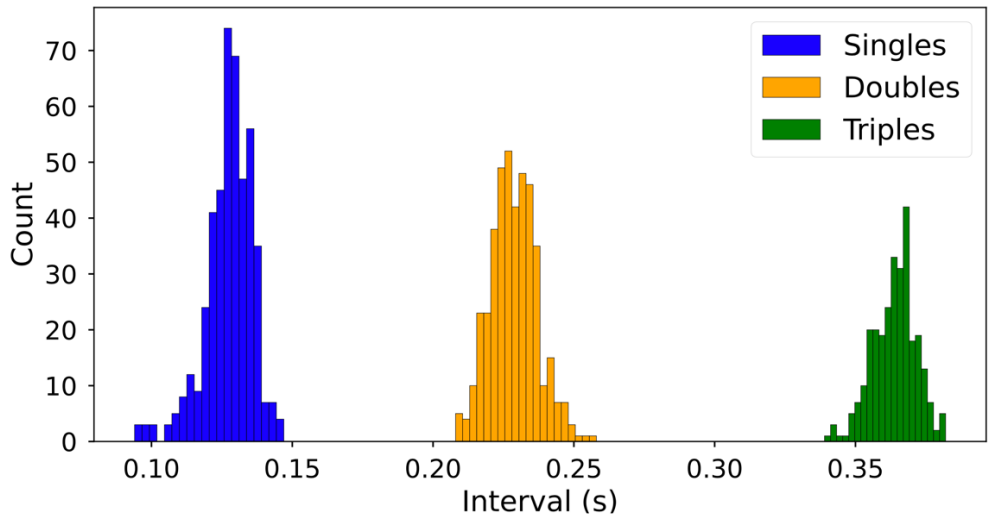

*Figure 7. Histograms of three types of intervals according to the notation in Fig. 5.*

**Swing Ratio**

The average swing ratio of the song can be determined according to Fig. 3 by dividing the mean interval of doubles by the mean interval of singles between the triplets, effectively excluding the intra-triplet singles. We find a swing ratio of 1.79 for Rosanna. Specifically, instead of the expected 1:2:3 ratios from musical notation, we find 1 : 1.792 : 2.852, and the double-triple ratio becomes 2 : 3.183 (instead of 2:3).

The computed swing factor of 1.792 significantly deviates from the theoretical swing ratio of 2 suggested by drum notation, indicating that the triplets are "suppressed" in time relative to their spacing. To our knowledge, swing ratios in shuffle grooves within pop and rock music have not been extensively studied. In contrast, research on jazz has shown that at slower tempos (under 150 bpm), swing ratios can reach values of 3 to 3.5, gradually decreasing to 1 (indicating no swing) as tempos approach 300 bpm (Friberg and Sundström, 2002). Comparing these findings to Rosanna, which features a half-time shuffle, is challenging. However, a swing ratio less than 2 suggests that Rosanna lacks an explicit swing feel in the traditional sense. This outcome is consistent with the characteristics of the Rosanna shuffle, which maintains a straightforward groove without aiming for a jazz-style swing.

**Tempo Drift**

Next, we examine the changes in tempo for "Rosanna". The upper panel of Fig. 8 displays the tempogram generated using Sonic Visualizer software (Cannam, 2010). We observe that the tempo fluctuates between 82-87 BPM, with these variations corresponding to different sections of the song depicted in the middle panel: A1-verse, A2-pre-chorus ("*Not quite a year since you went away...*") and B-chorus. Additionally, the synth and guitar solos, as well as the final jam, are also highlighted. The lower panel of Fig. 8 shows the *drift* of the eighth-note triplet pulse throughout the song. The gaps resulting from missed onsets are represented as gray areas, during which the drift is defined as zero. Also the changes in the drift align with the different sections of the song. This relationship between drift and song structure resembles findings reported for the 16th-note hi-hat pattern in "I Keep Forgettin'" (Räsänen, 2015). Recent studies have also investigated the physical characteristics of the drift, indicating that as a time series, it exhibits superdiffusive properties (Räsänen, 2024).

The distinction between the tempogram and the drift in Fig. 8 is due to the fact that the path is fundamentally a Brownian bridge, which may include a piecewise constant deterministic component (known as drift in stochastic processes). Additionally, it is not a stationary process, as its correlations are influenced by the number of onsets. This aspect related to stochastic modeling will be addressed in a separate study beyond the scope of the current work.

We also note that the drift in Fig. 8 clearly indicates that the song was recorded without a metronome or click track. The drift exceeds 1 second within the first 50 seconds, and such a deviation from a click track would not have been feasible, even though professional drummers can drift locally around the click. This conclusion aligns with the fact that Jeff Porcaro was not known to frequently use click tracks.

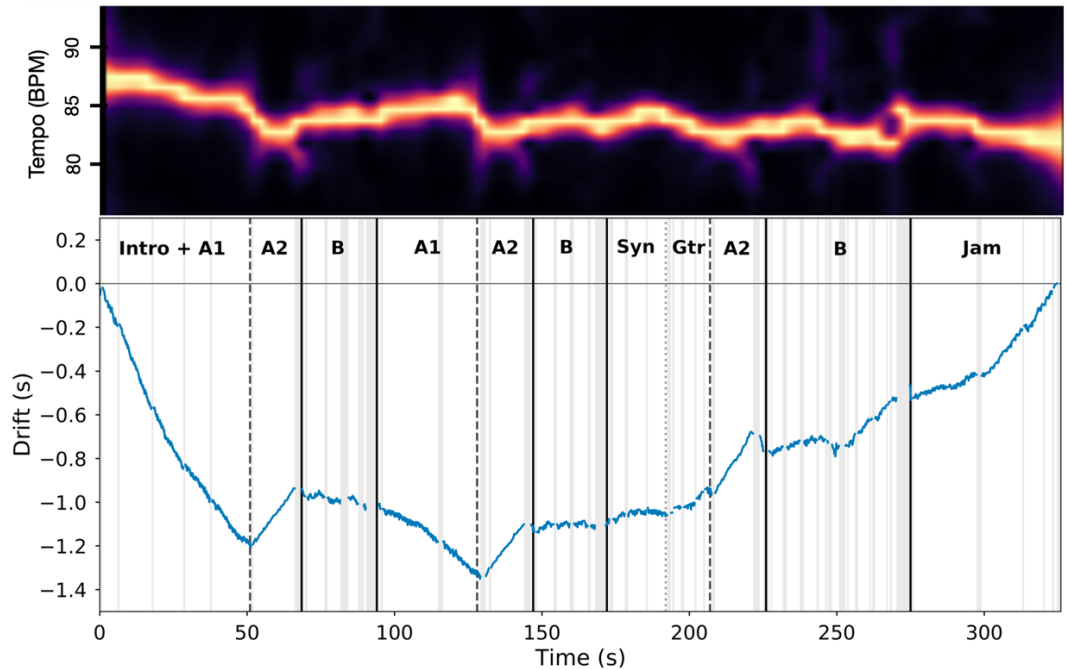

*Figure 8. Upper panel: Tempogram of "Rosanna" computed with Sonic Visualiser (Cannam, 2010); for details see the Methods section. Lower panel: Tempo drift during different sections of the song. A1: verse, A2: pre- chorus, and B: chorus.*

**Short-Scale Phrasing**

Next, we examine the short-scale interval and amplitude variations occurring within a two-bar phrase of the Rosanna shuffle. We included all two-bar phrases that had no missing onsets, excluding the snare ghost notes. A total of 12 phrases with complete onsets were identified.

Figure 9 displays the notation of the two-bar phrase along with percentages that indicate the relative differences between the averaged specific interval lengths and the average duration of the single eighth-note triplets calculated within the two-bar phrase. A negative value signifies that the interval is shorter than average, while a positive value indicates it is longer. The key observation is that all the longer intervals are shorter than average, reinforcing the swing ratio of less than two (1.79) analyzed earlier. Additionally, we note that the interval just before the snare is slightly prolonged compared to the other inter-triplet gaps. This can be interpreted as a laid-back feel typical in rock music, where the snare is slightly delayed in relation to the bass drum.

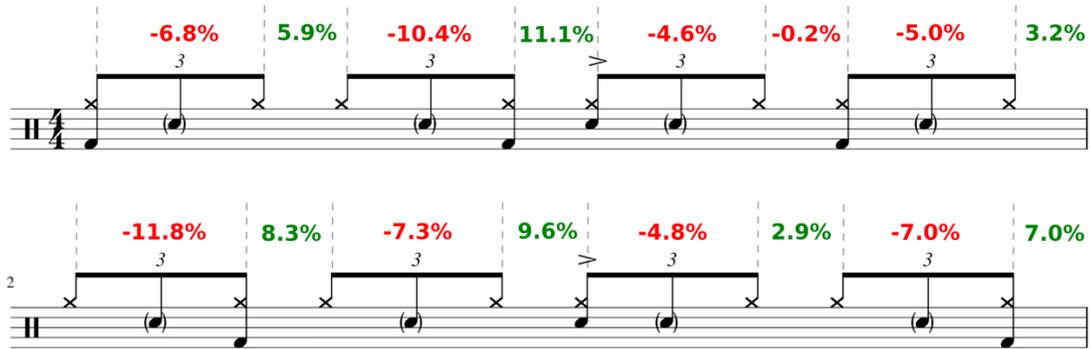

*Figure 9. Two-bar phrase along with percentages that indicate the relative differences between the averaged specific interval lengths and the average duration of the single eighth-note triplets calculated within the two-bar phrase.*

Figure 10 displays the mean amplitudes of the onsets that make up the shuffle, along with their respective standard deviations. In total, all 60 bars of the shuffle were analyzed, including the twelve-bar intro and both verses of the song. As mentioned earlier, some onsets could not be detected with sufficient accuracy, meaning that not all notes are represented in the data. However, there were enough onsets to draw meaningful conclusions from the analysis.

The bass-snare drum pattern is clear, as the amplitudes of hi-hats played alongside the bass or snare drum are significantly higher than those of other notes. While there is some leakage from the bass and snare to the hi-hat onset, it does not substantially affect the analysis. The last and first notes of various triplets, where only a hi-hat is present, reveal an intriguing pattern. The amplitudes form a consistent low-high sequence, with the lowest amplitude (excluding the ghost snare) corresponding to the last note of the triplet, while the first note has approximately double the amplitude. A similar pattern featuring a sub-structure in drumming dynamics has been observed in previous studies, where the hi-hat on the beat is distinctly accented (Räsänen, 2015).

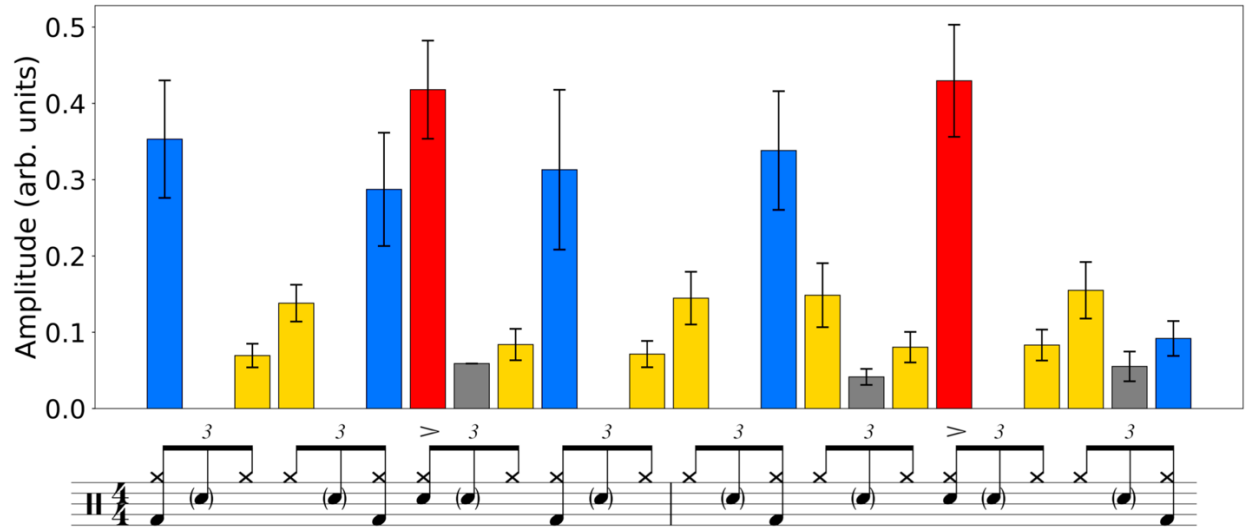

*Figure 10. Mean amplitudes of the onsets that make up the shuffle, along with their respective standard deviations.*

**Scaling Properties**

Finally, we apply detrended fluctuation analysis introduced above to the interbeat intervals and the onset amplitudes to examine their scaling properties and characteristics as complex time series. Figure 11 presents the DFA fluctuation functions F(s) as functions of the scale (s) for all intervals (a), singles (b), doubles (c), and triples (d). For all intervals, we find clear LRCs with at higher scales ($\alpha_2$ = 1.19), whereas the short-scale behavior is characterized by a scaling exponent typical for white noise ($\alpha_1$ = 0.58). The shift in behavior occurs at approximately 20-30 interbeat intervals, which corresponds to the two-bar phrase illustrated in Fig. 9. Therefore, we can anticipate that the anticorrelations in timing within the two-bar phrase contribute to a decrease in the scaling exponent in the short-scale regime.

Overall, this behavior is qualitatively similar to that observed in the straight 16th-note hi-hat pattern in "*I Keep Forgettin*'", which was analyzed previously (Räsänen, 2015).

The scaling properties of the singles and doubles are qualitatively similar to those of all intervals, with a higher scaling exponent at longer scales indicating long-range correlations (LRCs). Interestingly, the triples do not exhibit significant changes in the exponent; instead, the overall scaling exponent remains around $\alpha = 0.89$ across the scales, reflecting clear pink-noise characteristics. It is important to note that the triples correspond to the intervals in the pre-chorus section, which has a straight quarter-note feel, as illustrated in Fig. 2. Therefore, these sections of the song do not adhere to the two-bar phrasing that leads to short-scale anticorrelations for other intervals.

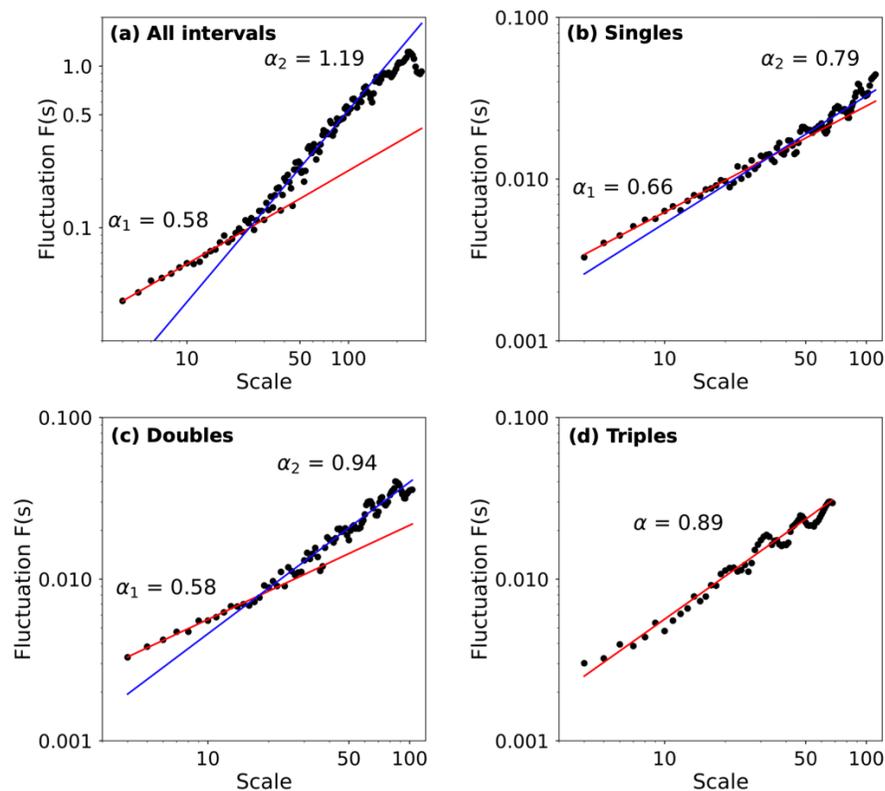

*Figure 11. Fluctuation functions of detrended fluctuation analysis (DFA) for all the interbeat intervals (a) and for the singles, doubles, and triples, respectively (b-d). The linear fits yield the DFA scaling exponents.*

In Fig. 12, the fluctuations of the onset amplitudes are illustrated. Similar to the intervals, we observe a change in slope at scales 20 to 30, which aligns with the two-bar phrase depicted for the amplitudes in Fig. 10. At smaller scales, the DFA exponent is $\alpha_1 = 0.53$, indicating white noise—resulting from a combination of intra-phrase dynamics in Fig. 10, which exhibit anticorrelated features, along with generic long-range correlation (LRC) characteristics in the dynamical variations. At longer scales, clear LRCs are observed, with a DFA exponent of $\alpha_2 = 1.02$, indicating 1/f fluctuations. Thus, this finding is consistent with previous studies (Räsänen, 2015).

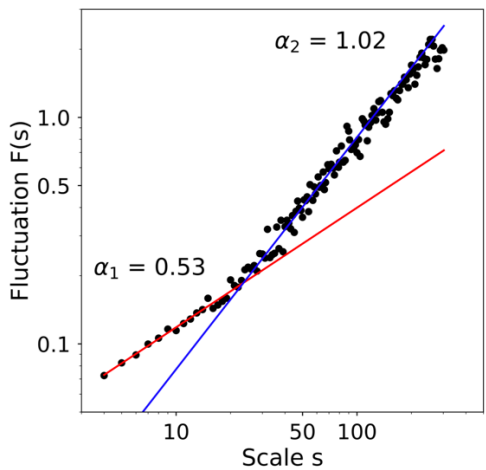

*Figure 12. Fluctuation function of detrended fluctuation analysis (DFA) for all the hi-hat onset amplitudes. The linear fits yield the DFA scaling exponents.*

## Discussion

Our results demonstrate a complex structure of timing and dynamic variations in the hi-hat pattern of the Rosanna shuffle. To our knowledge, shuffle structures have not been analyzed outside of a jazz context before. The observed swing ratio of 1.79 is noteworthy, as it significantly differs from the common notation for the song, which is typically represented as a triplet-feel shuffle with a swing ratio of 2:1. However, further studies are needed to compare this finding with other half-time shuffle grooves in popular music.

The cumulative interval fluctuations compared to the imaginary metronome revealed a clear drift, indicating that Jeff Porcaro recorded the track without a metronome. We also identified an intriguing structure and sub-structure within the two-bar phrase of the song, relating to both mean interval variations and amplitude variations. These nuances highlight a subtle phrasing in the hi-hat pattern of "Rosanna," which may contribute to the groove's significance in music history and drumming pedagogy. Lastly, we observed clear long-range correlations (LRCs) in both the fluctuations of interbeat intervals and amplitudes. These LRCs predominantly occur at larger time scales, while the short-range scaling behavior is influenced by the two-bar phrasing, which suppresses the LRCs in that range.

We identify several limitations in this work that warrant further investigation. First, we have examined only one sample of a single song; it would be valuable to analyze multiple recordings (including live performances) of the same song, played by the same drummer, Jeff Porcaro, as well as by other drummers who followed Porcaro in TOTO. Additionally, it is essential to explore other half-time shuffle grooves to uncover general and widely applicable features related to microtiming deviations in these rhythms, an area that is not well understood. In terms of the analysis itself, we focused here solely on data analysis using various methods and did not attempt to employ or develop models to comprehend the underlying stochastic processes. This will be addressed in future work. Nonetheless, we hope that this study inspires active further research in these areas.

## Conclusion

In this study, we have provided a detailed analysis of the timing and dynamics variations in Jeff Porcaro's iconic "Rosanna" shuffle. Our findings highlight the intricate rhythmic characteristics of this drum pattern, including a pronounced swing feel, long-range microtiming deviations, complex dynamical patterns, and tempo fluctuations that add to the track's distinctive groove. The two-bar structure and the interplay of hi-hat dynamics create a nuanced rhythmic flow, reinforcing the groove's impact on the song's phrasing. Overall, these results underscore the Rosanna shuffle's importance not just as a technical achievement, but as a prime example of rhythmic complexity in popular music. The combination of precision, feel, and subtle rhythmic variations solidifies its place as a benchmark in drumming and music production.